# Wideband Bandpass Filters Using a Novel Thick Metallization Technology

Celia Gomez-Molina*, Alejandro Pons-Abenza*, *Student Member, IEEE,* James Do,
Fernando Quesada-Pereira, *Member, IEEE,* Xiaoguang Liu, *Senior Member, IEEE,*
Juan Sebastian Gomez-Diaz, *Senior Member, IEEE,* and Alejandro Alvarez-Melcon, *Senior Member, IEEE*

*Abstract*—A new class of wideband bandpass filters based on using thick metallic bars as microwave resonators, instead of common microstrip lines, is presented. These bars provide a series of advantages over fully planar printed technologies, including higher coupling levels between resonators, better unloaded quality factors $Q_U$, and larger bandwidths, implemented with more compact structures. Moreover, thick bar resonators can easily be coupled to an additional resonance excited in a box used for shielding, allowing to realize transversal topologies able to implement transmission zeros at desired frequencies. To illustrate the capabilities of this technology, two microwave filters with different bandwidths and one transmission zero have been designed. One of the filters has been manufactured and tested using copper bars inside an aluminum housing partially filled with Teflon. Measured data demonstrates a fractional bandwidth about $FBW$ = 32%, spurious free range $SFR$ > 50%, unloaded quality factor of $Q_U$ = 1180 and return losses over 20 dB without requiring any post-tuning on the prototype, confirming the exciting performance of the proposed technology.

## I. Introduction

WIDEBAND devices are required in modern communication systems, including wireless communications, 5G, satellites, or in artificial intelligence platforms. In these applications, where the footprint and volume are critical, planar technologies, such as microstrip [1], substrate integrated waveguide (SIW) [2], [3] or coplanar [4], [5] are usually employed. Using these technologies, the level of losses is normally sacrificed in order to produce structures with reduced dimensions. Among the planar solutions, microstrip technology [1] is one of the most employed in practice, due to the compactness achieved and ease of fabrication. Unfortunately, the most important limitations faced by microstrip resonator filters [6] for these applications are normally small bandwidths and low unloaded quality factors ($Q_U$).

Some previous works have tried to implement broadband filters using microstrip line resonators. For instance, in [7] a compact bandpass filter was proposed using a microstrip resonant cell (CMRC). The maximum fractional bandwidth ($FBW$) achieved is approximately 9%. In [8], a microstrip bandpass filter is proposed using one miniaturized center-stubbed half-wavelength resonator and tapped-line input/output (I/O) ports. In this case, the bandwidth is around 300 MHz at 4.4 GHz, leading to a fractional bandwidth of $FBW$ = 6.8%. Another solution is reported in [9], where the authors use concentric open-loop resonators to reduce the size of the filter and achieve a $FBW \approx$ 10%. It is challenging to enhance the bandwidth of these microwave filters beyond these limits, mainly because very strong couplings are required between resonators and in the I/O ports. However, the coupling levels achievable with microstrip line resonators are low, which limit in practice the implementation of wideband responses. In spite of this, broadband planar filters are still possible to be implemented using non-resonant structures, such as using periodic arrangements [10], [11], or alternative configurations based on stepped-impedance stubs [12], [13].

The second limiting factor of printed microstrip technology is the relatively low $Q_U$. This factor is typically less than 150 for this technology. The low $Q_U$ is mainly due to the field distribution of the microstrip mode and the very small thickness of printed resonators. To overcome this limitation, the use of high dielectric constant strip resonators deposited on a low permittivity substrate was proposed in [14]. Using these dielectric resonators, filters with $Q_U \approx$ 880 were demonstrated, while the compactness is maintained. As observed, the achieved $Q_U$ using this technology is higher as compared to the traditional microstrip technology. However, in that work a standard tapped feeding line was employed to implement the I/O couplings. Due to this arrangement, rather low values could be implemented [14], resulting into a fractional bandwidth of only $FBW$ = 3.3%.

In applications where high $Q_U$ is strictly required, other non-planar alternatives, such as the waveguide technology [15], can be used. However, the compactness and low-cost manufacturing features of planar technology are compromised. Moreover, the achievable fractional bandwidths using waveguide resonators is typically less than 20%, as it was for instance reported in [15].

In this paper, we present a new approach to realize wideband

Manuscript received xxxx, 2019; revised xxxxx 27, 2019.

This work is supported in part by the National Science Foundation with CAREER Grant No. ECCS-1749177, and by the Spanish Government, Ministerio de Educación, Cultura y Deporte with Ref. FPU15/02883, and the Ministerio de Economía y Competitividad through the coordinated project with Ref. TEC2016-75934-C4-R and PRX18/00092.

*Both authors contributed equally to this manuscript.

C. Gomez-Molina, A. Pons-Abenza, F. Quesada-Pereira, A. Alvarez-Melcon are with the Department of Information Technologies and Communications, Universidad Politécnica de Cartagena, 30202 Cartagena, Spain. e-mail: {celia.gomez, alejandro.alvarez}@upct.es.

J. Do, X. Liu and J.S. Gomez-Diaz are with the Department of Electrical and Computer Engineering, University of California, Davis, USA. e-mail: {lxgliu, jsgomez}@ucdavis.edu

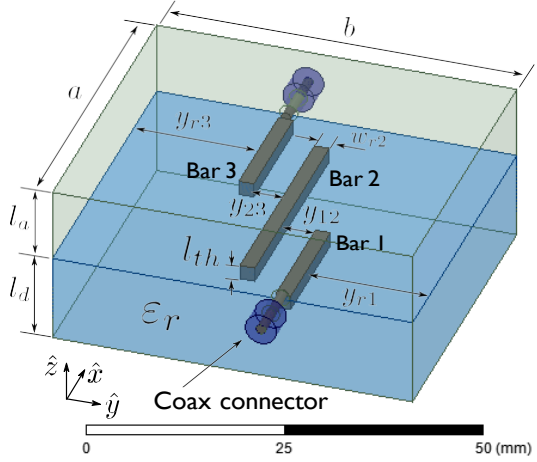

Fig. 1. 3D view of the proposed filter technology based on bulky bars buried in a dielectric medium. The sketch shows an example of a fourth order filter using three bar resonators coupled to the resonance provided by the house cavity. I/O couplings are implemented with the step impedance discontinuity produced from a direct connection of the coaxial connectors with the first/last bar resonators.

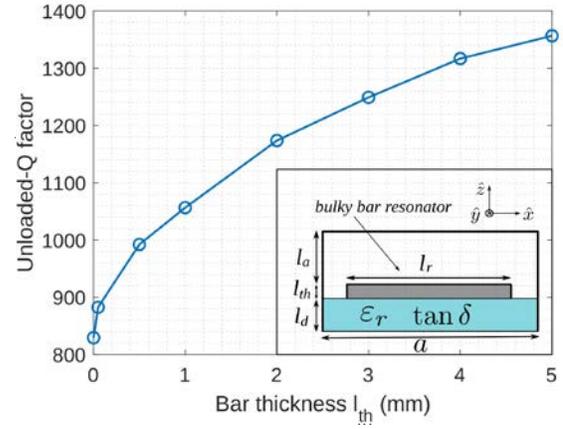

Fig. 2. Unloaded Q-factor ($Q_U$) of a single bulky bar resonator, as a function of the thickness ($l_{th}$). The geometry under test is shown in the inset. The board is Teflon with a relative permittivity of $\varepsilon_r = 2.1$ and a loss tangent $\tan \delta = 0.001$. The bar is made using copper with conductivity $\sigma_c = 5.8 \cdot 10^7$ S/m, and the box is made of aluminum with conductivity $\sigma_b = 3.8 \cdot 10^7$ S/m. Other dimensions are $l_d = 10$ mm, $l_a = 10$ mm, $l_r = 15$ mm, $w_r = 2$ mm (bar resonator width in the $\hat{y}$ axis), $a = 30$ mm and $b = 20$ mm (box dimension in the $\hat{y}$ axis). The test frequency is $f_c = 5.56$ GHz.

filters with reduced footprints, while maintaining large $Q_U$. A 3D view of the proposed technology is shown in Fig. 1. The proposed technology employs thick half-wavelength metallic bars buried in a dielectric medium instead of the typical printed metalizations used in microstrip technology.

In the following, we will demonstrate that the thick bars significantly increase the achievable coupling levels among resonators, while providing large $Q_U$. The use of metallic bars with large thickness also produces a strong decrease in the characteristic impedance of the guided *TEM* mode. Consequently, we demonstrate that I/O couplings can be implemented by introducing a step impedance discontinuity between the coaxial connectors and the first/last resonators. In this way, explicit I/O feeding lines, usually present in microstrip filters [1], are avoided, thus reducing the footprint of the structure as compared to standard microstrip printed technology. The adjustment of this step impedance discontinuity together with the capability to implement very large inter-resonator couplings, will enable to implement wideband frequency responses.

Similarly to microstrip technology, a metallic enclosure is employed to avoid spurious radiation (see Fig. 1). We show in this paper that an additional resonant mode can be excited in the shielding cavity partially filled with dielectric, similar to the idea proposed in [16], [17]. In those works, the I/O feeding lines were used to excite an additional cavity box resonance. In this new structure, some of the bar resonators, when adequately placed inside the cavity, can conveniently couple to this additional resonant mode. This will allow for the implementation of transversal topologies, which can be used to produce transmission zeros at finite frequencies in the insertion loss response of the filter.

In this work, two filters are designed using the proposed concept. A transversal-based topology of order four is implemented with three bar resonators and one box resonator, as shown in Fig. 1. The filters are designed at C-band, with center frequencies $f_{c1} = 3.26$ GHz and $f_{c2} = 3.35$ GHz, and bandwidths $BW_1 = 1.15$ GHz and $BW_2 = 1.575$ GHz, leading to fractional bandwidths of $FBW_1 = 35.28\%$ and $FBW_2 = 47.01\%$, respectively.

To validate the new concept, the first filter with lower bandwidth ($BW_1$) has been manufactured using thick bars made of copper, and Teflon as the dielectric material. The external housing is made from aluminum. Measurements obtained from the manufactured hardware show very good agreement against initial target specifications, and with respect to full-wave simulations obtained with Ansys High-Frequency Structure Simulator (HFSS) [18]. The results demonstrate that the technology proposed in this contribution represents a good trade-off between high $Q_U$ and compactness, allowing the implementation of very wideband responses, and with good spurious free range (*SFR*) performance.

## II. THICK BAR RESONATOR CHARACTERIZATION

The proposed technology is first characterized in terms of achievable $Q_U$, by using one isolated metallic bulky bar acting as a microwave resonator (see inset of Fig. 2). In contrast to printed microstrip resonators, the proposed technology uses bars with a non negligible thickness ($l_{th}$, see Fig. 1). In Fig. 2 we plot the value of $Q_U$ for different thickness of the resonator $l_{th}$. For this test, the bar thickness is varied in the range of $0.1$ mm $< l_{th} < 5$ mm. The bar resonator is made of copper with conductivity $\sigma_c = 5.8 \cdot 10^7$ S/m, and it is placed above a dielectric medium made of Teflon ($\varepsilon_r = 2.1$, $\tan \delta = 0.001$) with thickness $l_d = 10$ mm. All the elements are placed inside a cavity made of aluminum with conductivity $\sigma_b = 3.8 \cdot 10^7$ S/m.

According to Fig. 2, results show a strong dependence of $Q_U$ with the resonator thickness $l_{th}$. Maximum $Q_U$ of 1360 is obtained for the largest thickness $l_{th} = 5$ mm. This value drops to $Q_U = 830$ for the thinnest considered metalization of $l_{th} = 0.1$ mm. This study, therefore, shows the first benefit of the proposed technology, namely, the possibility of having significantly larger $Q_U$ values as compared to those obtained using printed resonators. This large $Q_U$ can be attributed to the extended surface provided by the bulky bars, through which the induced currents can flow.

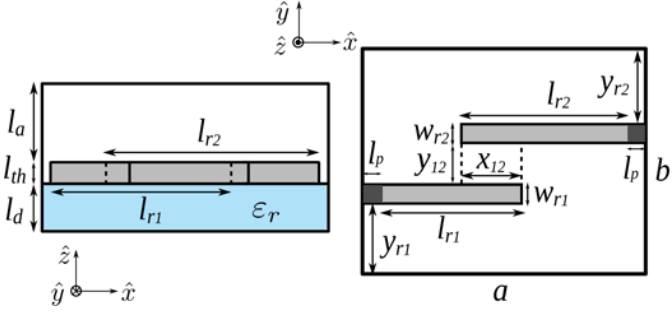

Fig. 3. Top (right panel) and side (left panel) view of two coupled thick bar resonators, showing relevant dimensions. Feeding ports are shown with dark areas $l_p$.

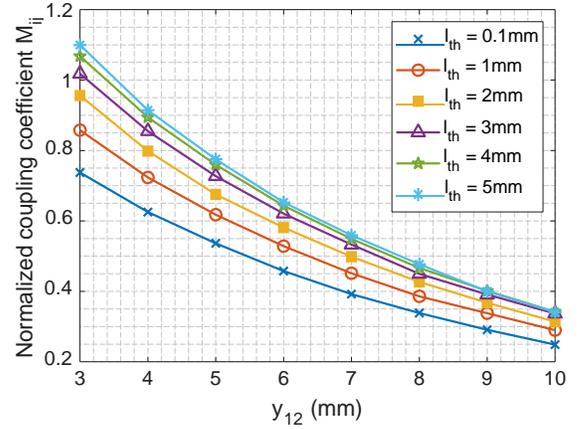

Fig. 4. Normalized coupling coefficient as a function of the separation between resonators $y_{12}$, for various bar thicknesses $l_{th}$. Coupling coefficient is normalized with the first frequency plan ($f_{c_1} = 3.26$ GHz, $BW_1 = 1.15$ GHz, $FBW_1 = 35.28\%$). Other dimensions are: $a = 30$ mm, $b = 28$ mm, $y_{r1} = y_{r2} = 10$ mm, $w_{r1} = w_{r2} = 2$ mm, $x_{12} = 0$ mm, $l_d = 10$ mm, $l_a = 10$ mm, $\varepsilon_r = 2.1$.

The next step is to characterize the inter-resonator coupling when the bulky bar technology is used. For this purpose, let us consider two symmetric coupled resonators as illustrated in Fig. 3. The dark areas in this figure $l_p$ represent the feeding ports of the resonators. With this setup, we calculate a normalized inter-resonator coupling coefficient $M_{ij}$ between the two resonators, using the described frequency plan for the bandwidth $BW_1$ [6]

$$M_{ij} = \frac{f_{c_1}}{BW_1} k_{ij} \qquad (1)$$

where $k_{ij}$ is the inter-resonator coupling coefficient extracted from the even/odd frequencies of the two coupled resonators. As usual, the test is done under very weak I/O couplings.

In Fig. 4, the normalized coupling coefficient $M_{ij}$ is shown as a function of the separation between bars $y_{12}$, and for different thicknesses $l_{th}$. Similarly to two coupled microstrip printed resonators, the inter-resonator coupling in this structure can be controlled using both $y_{12}$ and the overlapping distance along the $\hat{x}$ axis ($x_{12}$, see Fig. 3). This is observed in the plot, since the normalized coupling coefficient clearly increases as $y_{12}$ is decreased. What it is more interesting is the behavior of the inter-resonator coupling with the bar thickness. The plot clearly indicates that the inter-resonator coupling can be significantly increased using thick bars. In particular, if the thickness of the bar resonator is small ($l_{th} = 0.1$ mm), the maximum normalized coupling value is $M_{ij} = 0.78$. This coupling can be increased to 1.15 if $l_{th}$ is increased to 5 mm. This study shows the second benefit of the proposed technology, which is the possibility to obtain larger inter-resonator couplings as compared to the couplings that can be traditionally obtained with standard printed microstrip line resonators. The larger couplings can be attributed to a stronger capacitive effect that is formed, when the thickness of the bar is increased.

The implementation of wideband responses requires not only of large inter-resonator couplings, but also a mechanism to implement large I/O couplings. In this context, it is important to mention that another difference between the proposed technology and traditional microstrip implementations is the characteristic impedance of the transmission line ($TEM$) mode, formed by the thick bar structure. In general, microstrip filters are designed with standard coaxial transitions with $50\,\Omega$ or $75\,\Omega$ transmission line impedances. The microstrip line is usually dimensioned to match the coaxial impedance, in order to avoid unnecessary reflections between the first/last printed lines and the source/load. Since no impedance mismatch is introduced, the first/last metalizations in these structures act as the feeding lines for the printed coupled-line resonators of the filter [1]. However, when the thickness of the metalization increases, the characteristic impedance substantially drops. Consequently, large impedance mismatch can be introduced between the coaxial connectors and the thick bar acting as a transmission line.

The idea proposed in this contribution is to implement the I/O couplings of the filter by directly using this impedance mismatch. Following this idea, the feeding lines traditionally used in regular printed microstrip filters are eliminated, thus leading to more compact structures. The I/O couplings are simply implemented by the step impedance discontinuity introduced when directly connecting the pin of the coaxial connector to the first/last thick bar resonators (represented with dark areas $l_p$ in Fig. 3). Since no feeding lines are required, a more compact structure can be achieved with the proposed technology, as compared to a traditional printed microstrip filter. On this basis, compactness would be the third advantage of the proposed technology. In addition, the adoption of this strategy for the implementation of the ports, permits to obtain very large I/O couplings, as it will be shown next.

To illustrate the concept, a test with a singly loaded resonator is performed to obtain the external quality factor $Q_{ext}$ as a function of the characteristic impedance $Z_p$ of the input coaxial connector, modeled as a lumped port connected to the thick bar resonator. The structure employed for this test is depicted in the inset of Fig. 5. The test is performed by computing the maximum value of the group delay for the $S_{11}$ parameter in this singly terminated configuration [6]. Results are presented in Fig. 5 as a function of the port reference impedance $Z_p$ for different bar thickness $l_{th}$. The obtained values for small port impedances are around $Q_{ext} = 70 - 100$, while $Z_p > 10\,\Omega$ produce very low external quality factors ($Q_{ext} < 10$). We can also observe that when $Z_p$ is small, the use of thicker bars leads to smaller $Q_{ext}$. For instance, the largest $Q_{ext}$ is around 100 when $l_{th}$ is small ($l_{th} = 1$ mm). This value decreases to 70 if a thicker bar is used ($l_{th} = 5$ mm). However, the effect of the thickness in the $Q_{ext}$ is marginal when $Z_p > 10\,\Omega$.

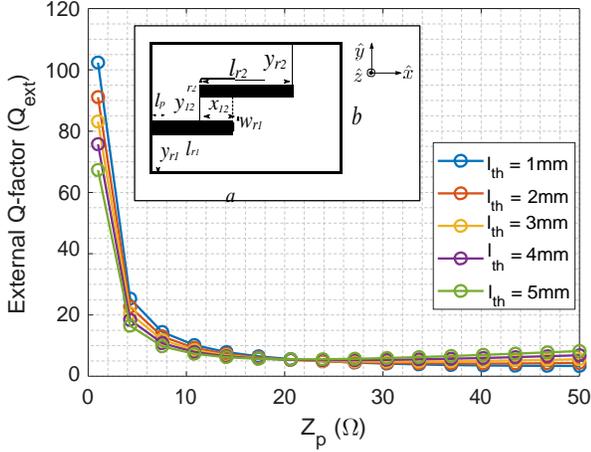

Fig. 5. External quality factor $Q_{ext}$ as a function of the input port reference impedance $Z_p$, for various values of the bar thickness $l_{th}$. The test frequency is $f_c = 3.5$ GHz. Other dimensions are: $l_d = 10$ mm, $l_a = 10$ mm, $w_r = 2$ mm, $a = 38$ mm, $b = 32$ mm. The dark area $l_p$ denotes a lumped port that models the direct connection of the coaxial pin to the first resonator.

The observed behavior indicates that very large I/O couplings are indeed achievable by properly adjusting the step impedance discontinuity introduced between the coaxial connector and the first/last resonator, and therefore there is no need to use extra I/O feeding lines. The opportunity to implement very large I/O couplings, finally demonstrates another benefit of the proposed technology, which is the possibility to synthesize very wideband responses.

## III. Design of Filters Based on Thick Bar Technology

To demonstrate the feasibility of the described technology, the design of two fourth order filters with two different bandwidths is addressed, according to the frequency plans specified in Section I. In addition, the first design ($f_{c1} = 3.26$ GHz, $BW_1 = 1.15$ GHz) has a prescribed transmission zero at $f_{z1} = 4.15$ GHz, while the second design ($f_{c2} = 3.35$ GHz, $BW_2 = 1.575$ GHz) has a prescribed transmission zero at $f_{z2} = 4.5$ GHz. The positions of the transmission zeros have been adjusted to achieve a ripple level of approximately 20 dB in the rejection band. Note that both filters have quite large fractional bandwidths ($FBW_1 = 35.28\%$ and $FBW_2 = 47.01\%$), which cannot be achieved with traditional planar technology based on resonant elements. These large bandwidths are also difficult to be implemented even with waveguide resonators.

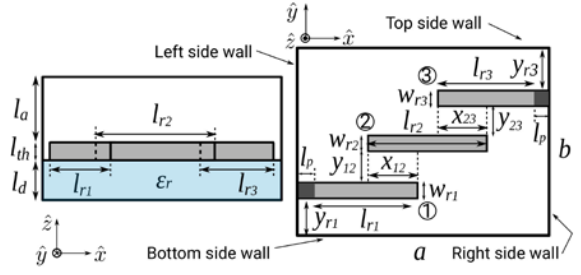

Fig. 6. Top (right panel) and side (left panel) view of the fourth order filters designed using the new proposed technology. Relevant dimensions are also shown. Thick bars are identified with numbers according to resonators shown in the coupling topology of Fig. 7. Dark areas ($l_p$) represent lumped ports that model the direct connection of the coaxial pins to the first/last resonators.

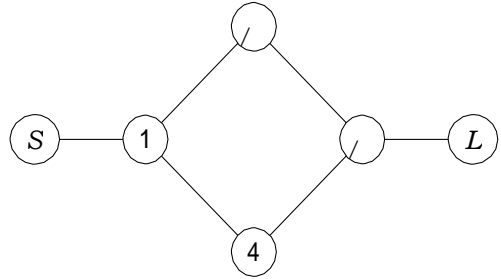

Fig. 7. Coupling topology that represents the filter structure shown in Fig. 6, with an additional cavity resonance. Bar resonators correspond to resonators 1, 2, and 3 in this topology, while the cavity resonance is represented by resonator 4.

The structure to be designed can be seen in Fig. 1. It will incorporate three thick bar resonators with the geometry sketched in Fig. 6. It should be remarked that the previously described mechanism for the I/O ports is used in this design. This is the reason why we cannot see extra I/O feeding lines in the structure shown in Fig. 6.

For the practical realization of the structure, the three thick bar resonators are buried in a dielectric material of height $l_d$. In addition to serving as supporting material for the thick bar resonators, the dielectric material partially filling the cavity allows to excite an additional box resonance, as described in [16], [17]. This box resonance will form the fourth resonator of our fourth order filter structure. The first/last bar resonators (bars 1 and 3 in Fig. 6) placed close to the left/right side walls of the cavity will couple to this resonance. On the contrary, the thick bar placed at the center (bar 2 in Fig. 6), will barely couple to this mode [16]. As a result, the coupling topology implemented with this arrangement is of the type known as transversal [19], and it is illustrated in Fig. 7.

The first three resonators in the coupling scheme represent the bar resonators of the physical structure, while the fourth resonator is implemented by the box mode of the cavity. The first and the third bars are coupled by proximity to the second bar resonator placed at the center of the box (bar 2 in Fig. 6), but they are also coupled to the cavity resonance (resonator number 4 in the coupling scheme of Fig. 7). This coupling topology can implement one transmission zero at finite frequencies [19].

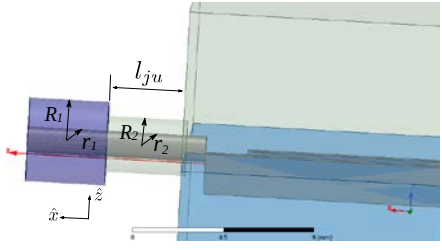

Fig. 8. 3D view of the coaxial transition employed to adjust the I/O couplings.

The $(N+2)$ coupling matrix of the designed filters for a transfer function with return losses $RL = 20$ dB is [6]

$$M_1 = \begin{bmatrix} 0 & 1.03 & 0 & 0 & 0 & 0 \\ 1.03 & 0.07 & 0.816 & 0 & -0.402 & 0 \\ 0 & 0.816 & 0.378 & 0.816 & 0 & 0 \\ 0 & 0 & 0.816 & 0.07 & 0.402 & 1.03 \\ 0 & -0.402 & 0 & 0.402 & -0.949 & 0 \\ 0 & 0 & 0 & 1.03 & 0 & 0 \end{bmatrix} \quad (2)$$

From this matrix, the first thing that is interesting to observe is that one of the four inter-resonator couplings must have negative sign. This sign change is important for implementing single-band responses. In the matrix shown, this is given by the elements $M_{14} = -M_{34}$, representing the coupling from bars 1 and 3 to the cavity resonance (resonator 4). As discussed in [16], this sign change is produced by a direction inversion in the electric field of the resonant mode when going from the left to the right side walls of the cavity.

The second important point is that, for achieving wide bandwidths, large I/O normalized couplings are required ($M_{S1} = M_{3L} = 1.03$). This challenge can be overcome by properly adjusting the step impedance discontinuity between the coaxial pin and the first/last bar resonators to obtain very small $Q_{ext}$. The external quality factors needed in our designs are calculated with [6]

$$Q_{ext_i} = \frac{f_{c_i}}{M_{S1}^2 \, BW_i} \quad (3)$$

leading to $Q_{ext_1} = 2.672$ and $Q_{ext_2} = 2.005$, respectively. The plot shown in Fig. 5 indicates that these are about the minimum values that can be obtained with the proposed I/O coupling mechanism.

For the practical realization of this concept, we propose to introduce a step discontinuity in the radius of a standard coaxial connector. This allows the employment of a commercial I/O connector, and at the same time permitting the implementation of the required $Q_{ext}$ values. Details of the arrangement are shown in Fig. 8. The nominal radii of the inner and outer conductors of the connector are $R_1$ and $r_1$ respectively, giving a characteristic impedance of $50\,\Omega$. Next, a length $l_{ju}$ is reserved in the cavity wall to introduce a step discontinuity in these radii. The internal conductor radius is provided by the commercial connector, and therefore it is easier to leave it untouched ($r_2 = r_1$). In order to lower the characteristic impedance of the coax, a step discontinuity is introduced in the radius of the outer conductor $R_2$. By selecting $R_2 < R_1$ and $r_2 = r_1$ (see Fig. 8), the characteristic impedance of the corresponding section can be substantially lowered.

For the implementation of the required I/O couplings, the commercial connector PE4128 from PASTERNACK was selected for the external section shown in Fig. 8, leading to outer and inner conductors radii of $R_1 = 2.1$ mm, $r_1 = 0.65$ mm. As mentioned before, the internal section that implements the step discontinuity takes the same radius for the inner conductor ($r_2 = r_1$), while it changes the radius of the outer conductor to $R_2 = 1.2$ mm. The second design follows the same strategy. However, since the bandwidth is larger, the step discontinuity needs to be smaller so then $R_2 = 1.4$ mm is used. These two examples show the high flexibility of the proposed technique to obtain very large I/O couplings.

Once the large I/O couplings are implemented, the next challenge is related to the fact that the required normalized coupling values between bars are also quite large ($M_{12} = M_{23} = 0.816$). This can be overcome by increasing the thickness of the bars, as detailed in the previous section. Once a proper thickness value is selected for the bar resonators, the inter-resonator coupling can be fine adjusted by acting on the variables $x_{12} = x_{23}$ and $y_{12} = y_{23}$ of Fig. 6. This is easily accomplished by producing charts similar to the one shown in Fig. 4.

The last coupling to be adjusted in order to implement the matrix shown in (2) is given by the element $M_{14} = -M_{34}$. This value represents the coupling between the first/last (1,3) bars and the cavity resonance. The coupling will be dependent on the distance from these bars to the top and bottom sidewalls (parameter $y_{r1} = y_{r3}$ in Fig. 6), as well as on the air and dielectric heights (parameters $l_a$ and $l_d$ in Fig. 6). Since the $\hat{x}$-component of the electric field is zero at the top/bottom side walls, this coupling increases when $y_{r1}$ and $y_{r3}$ increase. Certainly, it is much more difficult to implement strong couplings for this element by acting on all these geometrical variables. Fortunately, the required coupling in normalized terms ($M_{14} = -0.402$) is much smaller than for the previous couplings. This fact allows for the possibility to still implement passband responses exhibiting wide bandwidths. The second filter requires stronger coupling since it has a wider bandwidth. This is achieved by increasing $y_{r1} = y_{r3}$. However, this distance also affects the width of the box $b$, therefore modifying the resonant frequency of the cavity mode ($M_{44}$ of the coupling matrix). To correct for this shift of the resonant frequency, the overlapping distance between bars $x_{12} = x_{23}$ can be shortened, thus effectively decreasing the length of the box $a$. Alternatively, the resonant frequency of the cavity mode can also be adjusted by modifying the air and/or dielectric heights, $l_a$ and $l_d$. We note that the modification of the overlapping distance $x_{12} = x_{23}$ will, in turn, change the couplings between bars $M_{12} = M_{24}$. However, the effect is compensated, once more, with the adjustment of the separation between them $y_{12} = y_{23}$.

After all couplings and resonant frequencies are adjusted following this procedure, a global optimization is applied to the whole structure for fine tuning. The final dimensions of the two designed filters are included in Table I.

TABLE I
PHYSICAL DIMENSIONS OF THE TWO DESIGNED FILTERS, ACCORDING TO THE SKETCHES SHOWN IN FIG. 6 AND FIG. 8.

| Parameters of the filter | Dimensions (in mm) First design ($BW_1 = 1.15$ GHz) | Dimensions (in mm) Second design ($BW_2 = 1.57$ GHz) |
| --- | --- | --- |
| $l_a$ | 10 | 8 |
| $l_d$ | 10 | 8 |
| $l_{th}$ | 2 | 2 |
| $l_{r1} = l_{r3}$ | 16.1 | 15.98 |
| $l_{r2}$ | 30.2 | 26.6 |
| $w_{r1} = w_{r2} = w_{r3}$ | 2 | 2 |
| $y_{r1} = y_{r3}$ | 16.8 | 19.3 |
| $y_{12} = y_{23}$ | 4.5 | 3.2 |
| $x_{12} = x_{23}$ | 11.1 | 13.9 |
| $R_1$ | 2.1 | 2.1 |
| $R_2$ | 1.2 | 1.4 |
| $r_1 = r_2$ | 0.65 | 0.65 |
| $l_{ju}$ | 4 | 4 |
| $a$ | 42.2 | 32.76 |
| $b$ | 48.6 | 51 |

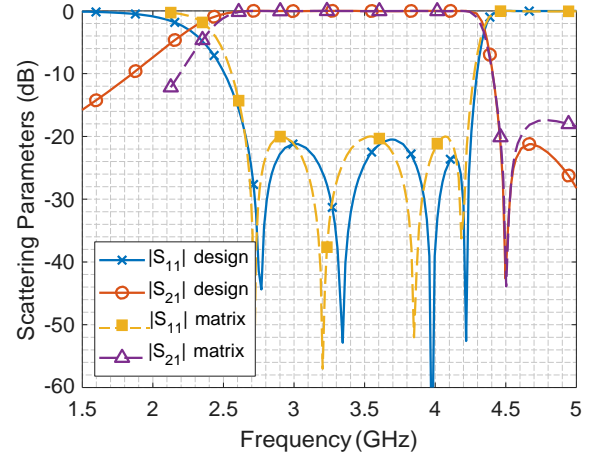

Fig. 10. Scattering parameters of the second designed filter obtained with HFSS (design) and comparison to the coupling matrix response (matrix). The bandwidth is $BW = 1.57$ GHz ($FBW = 47.01\%$). The dimensions of this second design are reported in the third column of Table I.

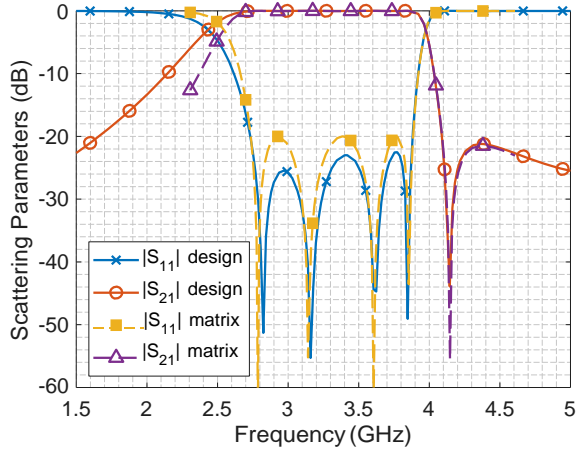

Fig. 9. Scattering parameters of the first designed filter obtained with HFSS (design) and comparison to the coupling matrix response (matrix). The bandwidth is $BW = 1.15$ GHz ($FBW = 35.28\%$). The dimensions of this first design are reported in the second column of Table I.

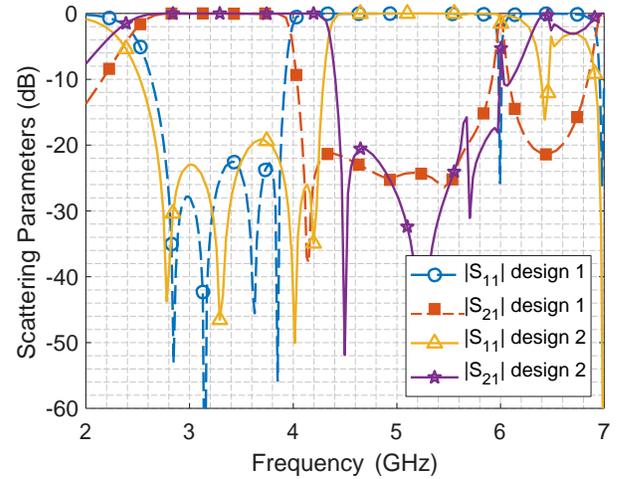

Fig. 11. Comparison for the out of band responses of the two designed filters from HFSS simulations. Dimensions of both filters are collected in Table I.

Details of the in-band responses for the two designed filters are shown in Fig. 9 and Fig. 10. Comparisons with respect to the responses of the coupling matrix are also included, showing that the design process was effective for both filters. In the two examples the transmission zero is placed approximately at the same distance from the passband edge, thus demonstrating that the passband characteristics can be controlled with the proposed structure, independently from the position of the transmission zero.

To conclude, a comparison between the two filters in terms of spurious free range ($SFR$) has been performed. The out of band responses of both filters are compared in Fig. 11. As it can be observed, the $SFR$ of the two filters is similar. In both cases, the first spurious band appears at $f_{s1} = 6$ GHz, and is due to the excitation of higher order resonances in the shielding cavity. This gives $SFR_1 \approx 2.1$ GHz (54 %) for the first filter and $SFR_2 \approx 1.8$ GHz (43 %) for the second design. This performance is far superior as compared to other transversal filters implementations previously reported. See for instance the transversal filter implemented in waveguide technology with $SFR = 21\%$, reported in [20].

## IV. PROTOTYPE MANUFACTURING AND MEASUREMENTS

A photograph of the manufactured prototype is presented in Fig. 12. Details of one of the copper thick bars, before assembling, can be observed in Fig. 12(a). As shown, the thick bars are buried into slots mechanized in the Teflon substrate. The aluminium enclosure used for shielding is manufactured in two pieces that are aligned using four dowel pins (see Fig. 12(a)). This allows to easily mechanize in the left/right side walls of the cavity the step discontinuities required in the radius of the outer conductor for the I/O coupling adjustment. A detail of this process in shown in the inset of Fig. 12(a). Details of the assembling process of the connector pins with the copper bars are shown in the inset of Fig. 12(b). The two pieces of the enclosure are finally assembled with the use of 24 pressure screws. Rounded corners of 2 mm are introduced in the different parts of the structure, as required by the milling process. They have shown to have little effect on the electrical behavior of the filter.

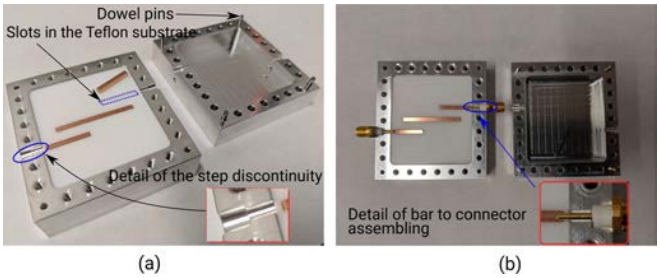

Fig. 12. Manufactured prototype. (a) Components of the filter before assembling, and details of the slots in the Teflon substrate used for alignment. The inset inside shows the step discontinuity needed in the radius of the outer conductor for the I/O coupling adjustment. (b) Top view of the structure after assembling. Inset shows details of the assembling process between the coax center pin and the copper bar.

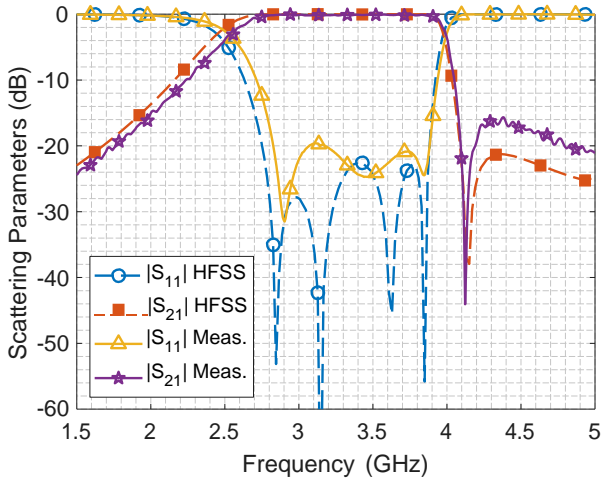

Fig. 13. Comparison between the S-parameters simulated using the full-wave commercial software HFSS and measured data from the manufactured prototype shown in Fig. 12.

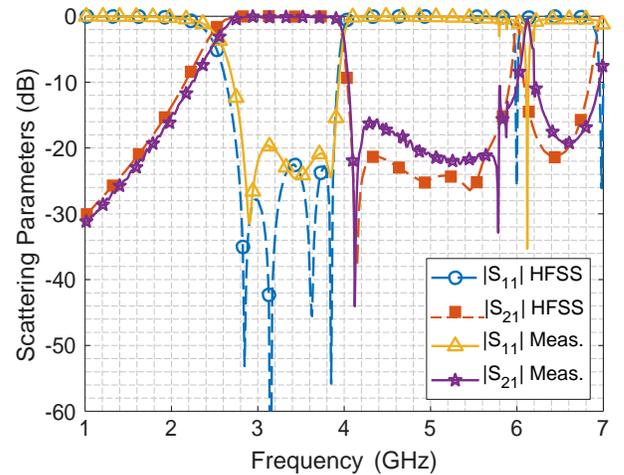

Fig. 14. Comparison between the S-parameters simulated using the full-wave commercial software HFSS and measured data from the manufactured prototype shown in Fig. 12 (out of band performance).

Measurements corresponding to the manufactured prototype are compared to the in-band full-wave response obtained using HFSS in Fig. 13. As can be observed, the in-band measured response shows a very good agreement with respect to the full-wave simulation, thus fully validating the proposed concept.

The measured bandwidth is $BW = 1.06$ GHz being slightly smaller than the target value. The transmission zero is slightly closer to the passband than expected. This raises the stopband ripple from 20 dB to 16 dB. Moreover, the return loss level is better than 20 dB. We note that two poles of the transfer function are complex due to manufacturing errors and imprecisions during the assembling process. It should be emphasized that the presented results are obtained directly from the manufactured prototype without applying any tuning operation.

The measured insertion losses are $IL = 0.16$ dB. This corresponds to an approximate value for the unloaded quality factor of $Q_U = 1180$. It is instructive to compare this value with those traditionally obtained using other compact technologies. The obtained $Q_U$ value is greater than values typically reported for microstrip technology ($Q_U < 150$) and even SIW technology ($Q_U < 640$) [2], [3]. In terms of $Q_U$, the performance of this filter is also better than the filter based on dielectric resonators proposed in [14], where a $Q_U = 880$ was reported.

Finally, we check the measured out of band response for the manufactured prototype and compare it to full-wave simulations in Fig. 14. Good agreement between measured results and the ideal response is also achieved in this case. These results experimentally confirm that the first spurious band appears at $f_{s1} = 6$ GHz, confirming a $SFR \approx 2.1$ GHz (54 %).

## V. CONCLUSION

In this contribution, a novel technology for the implementation of wideband filters has been presented and experimentally validated. The advantages of thick metallic bars as microwave resonators instead of common microstrip lines have been verified through various examples. Such advantages include the possibility to implement wide passbands, enabled by the high couplings between thick bars. Also, input/output feeding lines are avoided by exploiting the strong mismatch between bar resonators and standard coaxial connectors. This leads to more compact structures, and allows for the implementation of very strong I/O couplings, which is also required for implementing wideband responses. Moreover, the proposed bulky resonators exhibit higher unloaded quality factors $Q_U$ than other compact technologies such as microstrip or SIW. Finally, we have demonstrated that the suitable excitation of an additional box resonance allows for the implementation of transversal coupling topologies. In addition to the theoretical design process, a fourth order filter with fractional bandwidth of $FBW = 35.28\%$ and $Q_U = 1180$, has been experimentally demonstrated. Measured return losses of 20 dB and $SFR > 50\%$ were obtained. Results between the measured and simulated models show good agreement. The reported data experimentally confirm the performance of the proposed technology, and its promising future in wideband communication systems.